\documentclass[aps, prb, superscriptaddress, reprint, nobibnotes, twocolumn]{revtex4-2}
\usepackage{amsmath,amssymb}
\usepackage{color}
\usepackage{comment}
\usepackage{bm,graphicx,hyperref}
\hypersetup{%
  breaklinks = {true},
  citecolor = {blue},
  colorlinks = {true},
  linkcolor = {red},}

\usepackage{lineno}
\begin{document}

\title{On activation in solid ionic electrolytes}

\author{K. Noori}
\affiliation{NUS Institute for Functional Intelligent Materials, 4 Science Drive 2, Singapore 117544, Singapore}

\author{B. A. Olsen}
\affiliation{Department of Physics, Lewis \& Clark College, Portland, Oregon, 97219, USA}

\author{A. Rodin}
\affiliation{Yale-NUS College, 16 College Avenue West, 138527, Singapore}
\affiliation{Centre for Advanced 2D Materials, National University of Singapore, 117546, Singapore}
\affiliation{Department of Materials Science and Engineering, National University of Singapore, 117575, Singapore}

\begin{abstract}

Ionic conductivity in solid electrolytes is commonly expected to exhibit Arrhenius dependence on temperature, determined by a well-defined activation energy.
Consequently, a standard approach involves calculating this energy using quasi-static methods and using the Arrhenius form to extrapolate the numerical results from one temperature range to another.
Despite the ubiquity of this Arrhenius-based modeling, disagreements frequently arise between theory and experiment, and even between different theoretical studies.
By considering a tractable minimal model, we elucidate the reason behind the breakdown of the Arrhenius conductivity form.
This breakdown is driven by non-trivial phase-space boundaries between conducting and non-conducting regimes, and depends on the kinetic properties of the system.

\end{abstract}	

\maketitle

\section{Introduction}
\label{sec:Introduction}

Ionic motion through solid electrolytes~\cite{MahanRoth, Mehrer} plays a central role in the operation of advanced batteries~\cite{Bachman2016, Manthiram2017, Famprikis2019}. 
One of the most important quantities associated with the performance of these electrolytes is their conductivity, generally accepted to take the form of $\sigma = \sigma_0 e^{-E_a / k_BT} / T$,~\citep{MahanRoth, Bachman2016, Manthiram2017, Famprikis2019} where $T$ is the temperature, $k_B$ is the Boltzmann constant, and $E_a$ is referred to as the activation energy.
In this formulation, ionic conductivity in solids increases with temperature; the exponential term determines the number of charge carriers, while $T^{-1}$ arises from the temperature dependence of mobility.
This dependence of $\sigma$ on $T$ is confirmed experimentally, with the most explicit demonstration being a linear relation between $\log(\sigma T)$ and $1 / T$~\citep{Rogez2000, Wilkening2008, Chu2016, Deng2017, Krauskopf2017}, and others  finding a linear dependence of $\log(\sigma)$ on $1 / T$~\citep{Deng2015, Kato2016, Fall2019, Yashima2021, Hu2023}.

When current-carrying ions move through the solid electrolyte, they navigate a nontrivial potential landscape produced by the stationary ions of the solid.
Traversing this landscape requires overcoming local energy maxima, which is achievable only for sufficiently energetic mobile ions.
Assuming that the ionic energies are Boltzmann-distributed and the energy barrier is given by $E_a$, the number of ions in the system that can overcome the barrier at any given time is $\propto e^{-E_a / k_BT}$, giving rise to the Arrhenius dependence of $\sigma$.
This statement can be rephrased if the ionic motion is assumed to be ergodic: the probability that a given ion manages to overcome the energy barrier is $\propto e^{-E_a / k_BT}$.

This fact has implications for the interpretation of numerical studies on conductivity. For example, molecular dynamics (MD) simulations---especially in their \textit{ab initio} formulation (AIMD)---are often used to estimate $\sigma$.
Owing to the high computational cost of AIMD, simulations are typically restricted to a small set of temperatures; conductivity values at other temperatures are then extrapolated using the Arrhenius relation.
Another common approach is to use nudged elastic band (NEB) simulations to compute the migration energy barrier, $E_\mathrm{NEB}$, along the presumed ionic migration pathway. This value is assumed to be equivalent to $E_a$ and thus serves as a proxy for $\sigma$.
However, there is experimental evidence of a breakdown in the linear relation between $\log(\sigma)$ and $1 / T$ in common solid electrolyte materials \cite{Byeon2021}. Moreover, recent AIMD studies with denser temperature sampling over extended simulation times using machine-learning assisted potentials have  also shown a departure from Arrhenius scaling \cite{Qi2021}.

To understand this deviation, note that the Arrhenius model assumes that the energy barrier is independent of temperature.
While this assumption is generally true for electronic semiconductors, where the gap does not vary with temperature, thermal vibrations of the lattice in ionic conductors lead to significant fluctuations of the potential energy surface. These fluctuations depend on temperature and preclude the existence of a well-defined and temperature-independent activation energy.

In this work, we develop a minimal model to demonstrate how even the simplest ionic systems deviate from Arrhenius scaling.
In Sec.~\ref{sec:Theory}, we propose a theoretical explanation of why the Arrhenius behavior will generally not arise in ionic conductors over extended temperature ranges.
To substantiate this model, in Sec.~\ref{sec:Numerical_simulations}, we consider a numerically-tractable system and demonstrate the breakdown of the Arrhenius scaling in the presence of thermal fluctuations.
We conclude with a discussion of our results and their implications for state-of-the-art numerical techniques in Sec.~\ref{sec:Conclusions}.

We perform all our calculations using \texttt{JULIA}~\citep{Bezanson2017} and make our code available at https://github.com/rodin-physics/oscillator-activation.
Our plots are visualized using Makie.jl,~\citep{Danisch2021} employing a color scheme suitable for color-blind readers, developed in Ref.~\citep{Wong2011}.

\section{Theory}
\label{sec:Theory}

\begin{figure}
    \centering
    \includegraphics{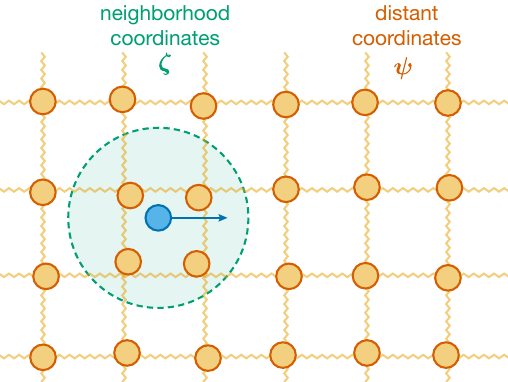}
    \caption{\emph{Mobile particle moving through a lattice.} Lattice ions sufficiently close to the mobile particle, included in the coordinate subset $\boldsymbol{\zeta}$, directly affect the motion of the particle. All other (distant) ions' coordinates are included in $\boldsymbol{\psi}$. The interaction between the distant ions and the mobile particle is sufficiently weak to be neglected when computing the particle's trajectory. 
    }
    \label{fig:Lattice}
\end{figure}

Consider a large system with a single mobile ion.
At a particular time, let the velocity and position of the ion be given by $\left(\mathbf{V},\mathbf{R}\right)$, six phase space coordinates in total.
Additionally, let the velocities and positions of all the other ions in the system be described by a $6N$-dimensional vector $\boldsymbol{\xi}$, where $N$ is the number of ions in the solid.
For any $\mathbf{R}$, the mobile ion lies within a polyhedron whose vertices correspond to the solid's atoms (see Fig.~\ref{fig:Lattice} for a 2D analog).
As the system evolves in time, the ion will either cross the face of the polyhedron toward which it is headed or be reflected.
Thus, in the phase space of $(\mathbf{V}, \mathbf{R},\boldsymbol{\xi})$, there are two groups of points: the ones which result in the ion traversing the polyhedron face and the ones that result in its reflection.
The escape probability is then given by the number of $(\mathbf{V}, \mathbf{R},\boldsymbol{\xi})$ escape configurations, weighted by their probability $\propto e^{-\beta E(\mathbf{V},\mathbf{R},\boldsymbol{\xi})}$, where $ E(\mathbf{V},\mathbf{R},\boldsymbol{\xi})$ is the energy of the configuration and $\beta = 1 / k_B T$.
We underscore that the energy of the entire system follows the Boltzmann distribution, as expected for a macroscopic system at some temperature $T$.
Defining the set of the escape configurations by $\mathcal{A}$ (for activation), we get the following expression for the escape probability

\begin{equation}
    \mathcal{P}
    =
    \left[\sum_{\mathbf{V},\mathbf{R},\boldsymbol{\xi}} e^{-\beta E(\mathbf{V},\mathbf{R},\boldsymbol{\xi})}\right]^{-1}
     \sum_{(\mathbf{V},\mathbf{R},\boldsymbol{\xi})\in \mathcal{A}} e^{-\beta E(\mathbf{V},\mathbf{R},\boldsymbol{\xi})}\,.
    \label{eqn:Escape_probability}
\end{equation}
From Eq.~\eqref{eqn:Escape_probability}, it is not immediately obvious that $\mathcal{P}\propto e^{-\beta E_a}$, as would be the case if the conductivity had the Arrhenius form.
However, for $N\rightarrow\infty$, it is difficult to provide a definitive proof one way or another.
Therefore, to make progress, we make a major simplification.

We assert that the escape success is determined by a subset of $\boldsymbol{\xi}$ coordinates, corresponding to the solid's ions sufficiently close to the mobile ion's position.
This assertion allows us to redefine $\boldsymbol{\xi} \rightarrow(\boldsymbol{\zeta},\boldsymbol{\psi})$, where $\boldsymbol{\zeta}$ are the coordinates that determine the escape and $\boldsymbol{\psi}$ are the remaining ones, as shown in Fig.~\ref{fig:Lattice}.
Moreover, we assume that the newly-defined $\boldsymbol{\zeta}$ couples weakly to $\boldsymbol{\psi}$, allowing us to neglect the influence of $\boldsymbol{\psi}$ on the time evolution of $\boldsymbol{\zeta}$ for sufficiently short times.
Consequently, $\boldsymbol{\psi}$ only plays a role in the probability of a configuration $(\mathbf{V}, \mathbf{R},\boldsymbol{\zeta})$, but not determining which configurations result in the ion's escape.
Hence, redefining $\mathcal{A}$, we have

\begin{align}
    \mathcal{P}
    &=
    \left[\sum_{\mathbf{V},\mathbf{R},\boldsymbol{\zeta}, \boldsymbol{\psi}} e^{-\beta E(\mathbf{V},\mathbf{R},\boldsymbol{\zeta}, \boldsymbol{\psi})}\right]^{-1}
     \sum_{(\mathbf{V},\mathbf{R},\boldsymbol{\zeta})\in \mathcal{A}, \boldsymbol{\psi}} e^{-\beta E(\mathbf{V},\mathbf{R},\boldsymbol{\zeta}, \boldsymbol{\psi})}
     \nonumber
     \\
     &=
     \sum_{(\mathbf{V},\mathbf{R},\boldsymbol{\zeta})\in \mathcal{A}}
     P(\mathbf{V},\mathbf{R},\boldsymbol{\zeta})
     \,,
    \label{eqn:Split_probability}
\end{align}
where we formally traced out $\boldsymbol{\psi}$ and defined the probability of the $(\mathbf{V},\mathbf{R},\boldsymbol{\zeta})$ configuration $P(\mathbf{V},\mathbf{R},\boldsymbol{\zeta})$.

\begin{figure}
    \centering
    \includegraphics{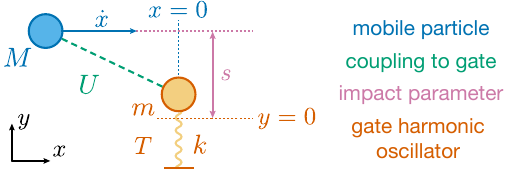}
    \caption{\emph{Single-mass lattice model.} A particle of mass $M$ is constrained to motion along $x$ and interacts via a coupling $U$ with a gate harmonic oscillator of mass $m$, which is constrained to motion along $y$. The axis of the mobile particle is separated from the equilibrium position of the oscillator by an impact parameter $s$. The gate oscillator has spring constant $k$, with mode occupation consistent with a temperature $T$.}
    \label{fig:Schematic}
\end{figure}

Importantly, $P(\mathbf{V},\mathbf{R},\boldsymbol{\zeta})$ cannot necessarily be assumed to have a Boltzmann dependence on the energy of the reduced system containing the coordinates $(\mathbf{V},\mathbf{R},\boldsymbol{\zeta})$.
Moreover, even if it does, $\mathcal{A}$ can take a complicated shape in the phase space so that the sum of $P(\mathbf{V},\mathbf{R},\boldsymbol{\zeta})$ over $\mathcal{A}$ should not be expected to give $e^{-\beta E_a}$, in general.
As an illustrative example, the following Section explores the simplest mobile ion-lattice configuration given in Fig.~\ref{fig:Schematic}.

\section{Numerical simulations}
\label{sec:Numerical_simulations}

\begin{figure*}
    \centering
    \includegraphics[width = \textwidth]{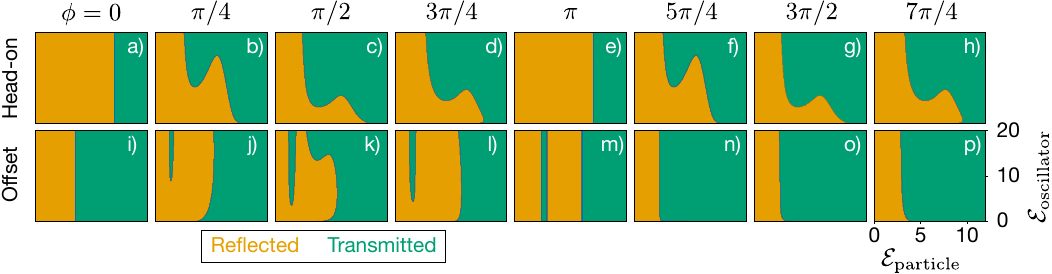}
    \caption{\emph{Phase space.} In a system with $M=1, \lambda = 2, \Phi_0=10$, initialized with various $\mathcal{E}_\mathrm{particle}$, $\mathcal{E}_\mathrm{oscillator}$, and $\phi$, the mobile particle passes the oscillator or is reflected.
    The plots show the domains where the particle is reflected (orange) and transmitted (green; this is the region $\mathcal A$).
    Each panel shows a portion of the $\mathcal{E}_\mathrm{particle}$-$\mathcal{E}_\mathrm{oscillator}$ space for a fixed $\phi$.
    The simulations are performed for $s = 0$ (head-on; a--h) and $s = 2$ (offset; i--p).}
    \label{fig:Phase_Space}
\end{figure*}

\subsection{Equations of motion}
\label{sec:Equations_of_motion}

Figure~\ref{fig:Schematic} shows a system where a particle of mass $M$ moves in one dimension perpendicular to a harmonic oscillator with mass $m$ and force constant $k$, interacting with it via a repulsive coupling $U$.
If the mobile particle is launched towards the oscillator from some initial position $x$ with some initial speed $\dot{x}$, it will either pass it or be reflected.
Using the notation from Sec.~\ref{sec:Theory}, $\boldsymbol{\zeta}$ is a two-component vector containing the displacement and speed of the oscillator, $y$ and $\dot{y}$ respectively.
Whether or not the particle passes the oscillator depends on the initial conditions of $x$, $\dot{x}$, $y$, and $\dot{y}$, partitioning the four-dimensional space $\left\{x, \dot{x}, y, \dot{y}\right\}$ into two regions, as discussed in Sec.~\ref{sec:Theory}.
To explore the geometry of this phase space and study the role of temperature in the transmission, we start by writing down the Lagrangian for the two-mass system:
\begin{equation}
    L = \frac{1}{2}M\dot{x}^2 + \frac{1}{2}m\dot{y}^2 - \frac{1}{2}ky^2 - U(x, y)\,,
    \label{eqn:Lagrangian}
\end{equation}
where the oscillator and the mobile particle are constrained to move in single, perpendicular directions.

From the oscillator frequency $\Omega = \sqrt{k / m}$, we define a time scale $t_\mathrm{osc} = 2\pi / \Omega$, a length scale $l_\mathrm{osc} = \sqrt{\hbar/m\Omega}$ (corresponding to the quantum oscillator length), and an energy scale $E_\mathrm{osc} = \hbar\Omega = k l_\mathrm{osc}^2$.
Redefining $x\rightarrow x l_\mathrm{osc}$, $y\rightarrow y l_\mathrm{osc}$, $M \rightarrow mM$, and $U\rightarrow U E_\mathrm{osc}$ then dividing Eq.~\eqref{eqn:Lagrangian} by $E_\mathrm{osc}$, we get a dimensionless Lagrangian

\begin{equation}
    L = \frac{1}{8\pi^2}M\dot{x}^2 + \frac{1}{8\pi^2}\dot{y}^2 - \frac{1}{2}y^2 -U(x, y)\,,
    \label{eqn:Lagrangian_unitless}
\end{equation}
where the time derivatives are taken with respect to $\tau =  t / t_\mathrm{osc}$.
The corresponding equations of motion are

\begin{align}
    \frac{M \ddot{x}}{4\pi^2}  &= -\frac{d}{dx}U(x,y) \,,
    \nonumber
    \\
   \frac{\ddot{y}}{4\pi^2} &= - y-\frac{d}{dy}U(x,y)\,.
   \label{eqn:EOM}
\end{align}

\subsection{Transmission phase space}
\label{sec:Transmission_phase_space}

For convenience, we express $y$ and $\dot{y}$ using the amplitude and the phase of the oscillator $y_0$ and $\phi$, respectively.
Although the phase space for the two-mass system is four-dimensional, if the mobile particle starts far enough from the oscillator so that the the interaction vanishes, modifying the initial separation is equivalent to changing $\phi$, reducing the effective dimensionality of the phase space.

To illustrate the shape of $\mathcal{A}$ containing all the transmitting configurations, we choose $U(z) = U_0 e^{-z^2 / 2\lambda^2}$ with $z = \sqrt{x^2 + (y - s)^2}$, where $s$ is the impact parameter, $\lambda = 2$, and $U_0 = 10$.
The motivation behind this choice of $U(z)$ is its simplicity: one can easily change the strength and extent of the interaction by adjusting $U_0$ and $\lambda$, as was done in other works demonstrating dissipation in similar systems~\citep{Rodin2022a, Mahalingam2023, Mahalingam2023a}.
We set $M = 1$ and use two different impact parameters ($s=0$ and $s=2$).
Next, we take a range of particle energies  $1 / 10 \leq \mathcal{E}_\mathrm{particle}\leq 12$, oscillator energies $0\leq \mathcal{E}_\mathrm{oscillator}\leq 20$, and oscillator phases $\phi\in [0, \pi/4, \dots 7\pi/4]$.
Here $\mathcal E_\mathrm{particle}= E_\mathrm{particle}/E_\mathrm{osc}$ and $\mathcal E_\mathrm{oscillator}= E_\mathrm{oscillator}/E_\mathrm{osc}$.
For each triplet $(\mathcal{E}_\mathrm{particle}, \mathcal{E}_\mathrm{oscillator}, \phi)$, we get the particle's initial speed, as well as the oscillator's initial speed and position.
Then, we launch the particle towards the oscillator starting at $x = -20$, evolving the system in time using the fifth order Runge-Kutta procedure with the time step $\delta\tau = 5\times 10^{-4}$.
At each time step, the velocity and the position of the particle are checked and the time evolution stops when either the velocity of the particle becomes negative (reflection) or the particle passes the oscillator (transmission). 

We plot the results in Fig.~\ref{fig:Phase_Space} as a collection of domain maps in the $\mathcal{E}_\mathrm{particle}$--$\mathcal{E}_\mathrm{oscillator}$ space for each of the $\phi$'s.
We see that the shape of $\mathcal{A}$ is highly nontrivial, even in this simple model.
To calculate the transmission probability, one needs to integrate $P(\dot{x}, y_0, \phi)$ over $\mathcal{A}$:

\begin{equation}
    \mathcal{P}
     =
     \sum_{(\dot{x}, x, \dot{y}, y)\in \mathcal{A}}
     P(\dot{x}, x, \dot{y}, y)
     \equiv
     \sum_{(\dot{x}, y_0, \phi)\in \mathcal{A}}
     P(\dot{x}, y_0, \phi)
     \,.
    \label{eqn:1D_probability}
\end{equation}
Given the shape of the phase space, even if we assume that the total system energy is Boltzmann-distributed with $P(\dot{x}, y_0, \phi)\propto e^{-\mathcal{E}/\omega_T}$ (with $\omega_T=k_BT/E_\mathrm{osc}$) so that constant-$\mathcal{E}$ surfaces form quarter-circles on the $\mathcal{E}_\mathrm{particle}$--$\mathcal{E}_\mathrm{oscillator}$ plane, it is unlikely that the integral will exhibit an Arrhenius behavior.
To be certain, we ``integrate" over the phase space by conducting a series of four numerical experiments, where we send the mobile particle towards the oscillator and calculate its transmission probability, making different assumptions about the system.

\subsection{Stationary oscillator}
\label{sec:Stationary_oscillator}

For the first two experiments, the oscillator starts from rest with zero displacement so that $y_0 = 0$, making $\phi$ irrelevant.
By eliminating two out of three variables in Eq.~\eqref{eqn:1D_probability}, we have $\mathcal{P} = \sum_{\dot{x}\in \mathcal{A}} P(\dot{x})$.
In this simple case, for a repulsive interaction, the particle must have a certain minimum speed so that $\mathcal{P} = \sum_{\dot{x} \geq \dot{x}_\mathrm{min}} P(\dot{x})$, corresponding to some minimum kinetic energy.
If the particle energy follows the Boltzmann distribution $\propto e^{-\mathcal{E}_\mathrm{particle} / \omega_T}$, the probability of transmission will take the Arrhenius form, given by $e^{-\mathcal{E}_\mathrm{min}/\omega_T}$.
As we show below, the value of $\mathcal{E}_\mathrm{min}$ is not always easily apparent.

\begin{figure}
    \centering
    \includegraphics[width = \columnwidth]{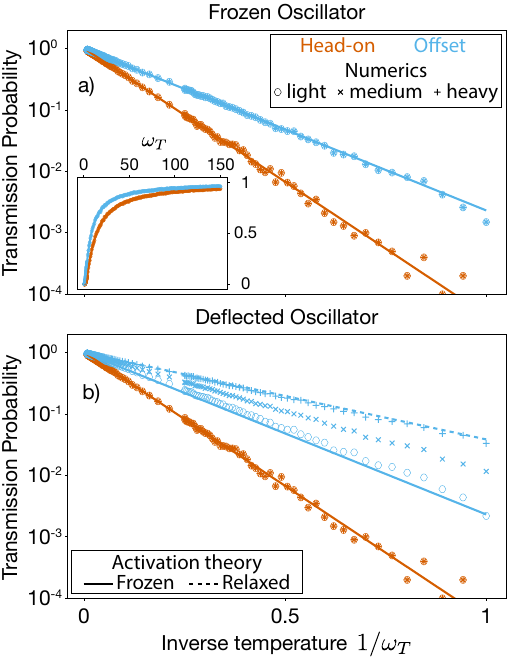}
    \caption{\emph{Stationary oscillator.} A particle starting from $x = -20$, is launched towards the oscillator, which is initially at rest, either head-on ($s = 0$) or offset to the side ($s = 2$).
    The speed of the particle is obtained by generating an energy $\mathcal{E}$ from $e^{-\mathcal{E}/\omega_T}$ and equating it to the kinetic energy $M\dot{x}^2 / 8\pi^2$ to calculate $\dot{x}$.
    Three masses are used ($1/5$, $1$, and $5$, referred to as ``light", ``medium", and ``heavy", respectively).
    The system is evolved in time until the particle either passes the oscillator or is reflected.
    For each mass at each temperature, 10,000 realizations are performed and the fraction of particles passing the oscillator is recorded.
    In panel (a), the oscillator is held fixed, resulting in a conservative potential for the particle.
    In panel (b), the system evolves in accordance with Eq.~\eqref{eqn:EOM}.
    Solid lines correspond to $e^{-\Phi(s) / \omega_T}$; the dashed line is given by $e^{-\mathcal{E}_* / \omega_T}$, where $\mathcal{E}_*$ is the minimized potential energy when the particle is at $x = 0$.
    }
    \label{fig:Stationary_Oscillator}
\end{figure}

If the oscillator is frozen in place and not allowed to move in response to its interaction with the particle, the minimum kinetic energy that the particle must start with to overcome the potential barrier is $\mathcal{E}_\mathrm{min} = U(s)$.
To establish the baseline, our first experiment aims to confirm this known result by considering a range of $\omega_T$'s and sampling 10,000 energies from the Boltzmann distribution at each $\omega_T$.
For each of the energies, the particle is launched towards the fixed oscillator from $x = -20$ and the system is evolved following the same Runge-Kutta procedure as in Sec.~\ref{sec:Transmission_phase_space}.
By counting how many of the 10,000 particles make it past the fixed oscillator at each $\omega_T$, we obtain the transmission probability as a function of temperature, which is plotted in Fig.~\ref{fig:Stationary_Oscillator}(a) for the same two impact parameters as in Sec.~\ref{sec:Transmission_phase_space}.
The experiment was repeated three times for different particle masses (1/5, 1, and 5) to confirm that the result is mass-independent, as expected.
By using the same random seed to generate the initial conditions for the three masses, we obtained identical results, as can be seen from the perfect overlapping of the scatter plot symbols.
The numerical results show an excellent agreement with the analytic Arrhenius form, validating our computation approach.

For the second experiment, we allow the oscillator to be deflected in response to the particle.
By performing the same steps as for Fig.~\ref{fig:Stationary_Oscillator}(a), we obtain the transmission probability and plot the results in Fig.~\ref{fig:Stationary_Oscillator}(b).
Because the oscillator can only move in the direction perpendicular to the particle's trajectory, for $s = 0$, there is no force component applied on the oscillator due to the vanishing derivative, so it remains stationary.
Comparing the results for $s = 0$ in panels (a) and (b) of Fig.~\ref{fig:Stationary_Oscillator}, we confirm that they are identical due to the same random seed.
For a finite impact parameter, on the other hand, the particle can push the mass laterally, reducing the maximum value of the interaction between the two objects and lowering the potential barrier.

In the spirit of NEB, one way to estimate the reduced barrier is to relax the system by bringing the particle to the point of the closest approach to the oscillator, minimizing the potential energy by varying $y$, and using this value for the barrier.
Explicitly, we minimize the potential energy by solving $U'(y_* - s)+y_*=0$.
In case there are multiple stable solutions, corresponding to the oscillator mass being on either side of the particle trajectory, we take the one that gives the smaller energy.
Using this minimized value as the activation energy $\mathcal{E}_\mathrm{min} =\mathcal{E}_*= \Phi(y_* - s) + y_*^2 / 2$ in the Arrhenius relation, we plot the ``relaxed'' theoretical curve in Fig.~\ref{fig:Stationary_Oscillator}(b) as a dashed line, along with the ``frozen'' result for an unrelaxed barrier $e^{-U(2) / \omega_T}$.

Comparing our numerical results for $s = 2$ to the Arrhenius plots with the two activation energies, we observe that the theoretical curves bound the numerical results.
Additionally, performing a power-law fit of $-\ln \mathcal{P}$ vs. $1/\omega_T$ for each mass's numerical data indicates that the dependence has an Arrhenius form.
This form indicates that there indeed exists an $\mathcal{E}_\mathrm{min}$, but it is mass-dependent and lies between the relaxed and unrelaxed values.
The mass dependence can be easily understood intuitively: for a given kinetic energy, heavier particles move slower compared to the light ones.
The slow speed allows the oscillator to move out of the particle's way and assume a configuration that is similar to the relaxed one.
Lighter particles, on the other hand, move faster and the oscillator does not have time to respond so that the particles need to overcome the unrelaxed (frozen oscillator) potential barrier.

There are two key points to take away from these experiments.
First, if only one of the system's degrees of freedom is probabilistic and follows the Boltzmann distribution, the transmission probability will take the Arrhenius form.
Second, one cannot generally determine the activation energy accurately while ignoring  kinetic effects, as can be seen from Fig.~\ref{fig:Stationary_Oscillator}(b).
Consequently, the activation energy obtained from quasi-static methods, such as NEB, is likely to disagree with experiment-derived values.
Next, we show that the existence of multiple probabilistic degrees of freedom, as would be the case in real systems, causes the transmission probability to deviate from the Arrhenius form.

\subsection{Moving oscillator}
\label{eqn:Moving_oscillator}

\begin{figure}
    \centering
    \includegraphics[width = \columnwidth]{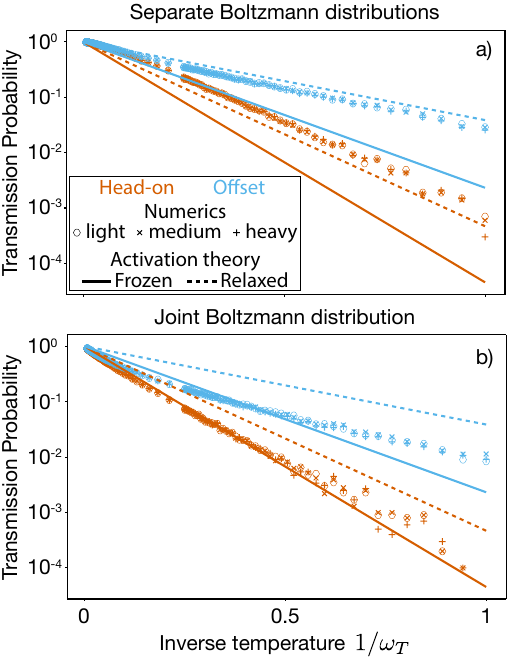}
    \caption{\emph{Moving oscillator.} Simulations following the same protocol as outlined in the caption of Fig.~\ref{fig:Stationary_Oscillator}, but with the oscillator not starting from rest.
    Instead, it is initialized with the initial position $\sqrt{2n+1}\sin\phi$ and velocity $2\pi\sqrt{2n+1}\cos\phi$.
    In panel (a), the probability distribution of the initial configurations is $ P(\dot{x}, y_0, \phi) =  P_\mathrm{particle}(\dot{x}) P_\mathrm{oscillator}(y_0)/2\pi$.
    In panel (b), the total energy of the system is distributed as $e^{-\mathcal{E}/\omega_T}$ with the fraction of energy distributed between the particle and the oscillator being random and uniform.
    In both cases, $\phi$ is uniformly distributed between $0$ and $2\pi$.}
    \label{fig:Moving_Oscillator}
\end{figure}

For the next two experiments, the oscillator is initialized with a finite speed and displacement for probabilistic $y_0$ and $\phi$.
In the first experiment, the particle and the oscillator are treated as two independent objects following independent Boltzmann distributions.
This scenario would arise if, prior to the particle being launched towards the oscillator, both objects were kept in contact with separate heat baths.
Following the same procedure as in Sec.~\ref{sec:Stationary_oscillator}, 10,000 realizations are performed at each $\omega_T$ with the particle energy sampled from the Boltzmann distribution $e^{-\mathcal{E}_\mathrm{particle} / \omega_T}$, while the oscillator energy is generated from $e^{-n/\omega_T}$ for integer $n$'s.
The phase $\phi$ is picked from a uniform distribution $[0, 2\pi]$.
The amplitude of the oscillator then becomes $\zeta = \sqrt{2n + 1}$ so that the initial position is $\zeta\sin\phi$ and the initial speed is $2\pi\zeta\cos\phi$.

The simulations are performed for the same masses and impact parameters used in Fig.~\ref{fig:Stationary_Oscillator}.
The results for the transmission probability are given in Fig.~\ref{fig:Moving_Oscillator}(a), along with the theoretical Arrhenius lines using relaxed and frozen activation energies.
Since the oscillator does not start from rest with $y_0 = 0$ and can be deflected by the particle even at  $s = 0$, the Arrhenius line for the relaxed configuration is also included.

The second experiment treats the particle-oscillator system as a single whole and assumes that the composite system's total energy is Boltzmann distributed.
To create such a configuration, one could imagine that the particle's motion is bounded by hard walls positioned far from the oscillator.
When the particle bounces off these walls, it exchanges energy with an external bath and later interacts with the oscillator so, after a long time, the two-body system acquires the temperature of the bath.
Here, the initial conditions for the 10,000 runs at each $\omega_T$ are generated by first picking the total energy $\mathcal{E}$ from the exponential distribution $e^{-\mathcal{E}/\omega_T}$.
After generating $\mathcal{E}$, we randomly pick $n$ between $0$ and $\lfloor \mathcal{E} \rfloor$ and set the energy of the particle to $\mathcal{E} - n$.
Finally, $\phi$ is chosen from a uniform distribution.
The results of this experiment are given in Fig.~\ref{fig:Moving_Oscillator}(b).

The two experiments represent maximally different scenarios: either the two degrees of freedom are completely independent or their composite system follows a thermal distribution.
In a real system, the situation is between the two extremes.
The statistics of the $\boldsymbol{\zeta}$ coordinates are not completely independent from the mobile particle.
At the same time, the mobile ion and its neighborhood do not comprise an isolated system with a Boltzmann energy distribution.
Figure~\ref{fig:Moving_Oscillator} shows that the Arrhenius formula does not capture the transmission probability for either case for both impact parameters, regardless of which activation energy is used.
We can see that, for either distribution, while the results for one impact parameter match a theory prediction reasonably well, the other impact parameter results clearly deviate from the Arrhenius form.
Notably, the impact parameter that is matched better by a theoretical result is different between the two panels, as is the method of obtaining the theoretical prediction (relaxed vs. frozen).
A closer look reveals that even the better-matched data exhibit a slight curvature, which would lead to a temperature-dependent activation energy.

The outcome of the experiments supports the discussion in Sec.~\ref{sec:Theory} that the transmission probability will, in general, not follow the Arrhenius form.
There are several implications stemming from this result.
First, extrapolating the results obtained using AIMD to different temperatures is precarious since the dependence of the activation energy on temperature is generally not known.
Moreover, even if one assumes that the activation energy is constant over a limited temperature range, it is not evident which temperature the NEB results should correspond to.
% In fact, as can be seen from Fig.~\ref{fig:Moving_Oscillator}, the NEB-like approach does a reasonable job for only one out of four configurations.

\section{Conclusions}
\label{sec:Conclusions}

In this work, we have explored the idea of activation in solid state ionic electrolytes.
By constructing a minimal model and visualizing the corresponding three-dimensional phase space, we have shown that the portion of this space that results in activation has a non-trivial shape.
Consequently, even if the probability of each point in the phase space follows the Boltzmann distribution, it is unlikely that summing over all the points in the activation region after weighting them with the appropriate Boltzmann factor will give rise to an Arrhenius dependence associated with activation.
To support this argument, we conducted a series of numerical experiments using our minimal model to demonstrate that one gets the Arrhenius curve only if there is a single Boltzmann-distributed degree of freedom and, at best, one can expect a slowly-varying temperature-dependent ``activation energy."
The form of this variable activation energy is non-trivial even for our simple configuration, making it difficult to reliably extrapolate the numerical results obtained at one temperature to different temperature ranges.
Finally, we have shown that one cannot accurately obtain the activation energy by neglecting the kinetic component of the ionic motion, as is commonly done in the quasi-static NEB calculations.

The strengths of existing numerical methods make them extremely useful
for rapidly screening large numbers of viable candidates for further experimental scrutiny.
While this work does not undermine the general usefulness of these methods, based on our findings, we suggest that caution must be taken when interpreting data from numerical simulations using the Arrhenius form of the $\sigma$--$T$ relation. Specifically, it is not guaranteed \textit{a priori} that such a relation exists for any given system, especially over a wide temperature range. 
Moreover, a single, temperature-independent activation energy for ionic migration is not likely to exist in general. 
This work points the way to areas of possible refinement for the theoretical simulation and study of ionic migration in solid electrolytes.

\acknowledgements

A.\,R. and K.\,N. acknowledge the National Research Foundation, Prime Minister Office, Singapore, under its Medium Sized Centre Programme. A.\,R. acknowledges support by Yale-NUS College (through Start-up Grant).
B.\,A.\,O. acknowledges support from the M.\,J. Murdock Charitable Trust.

% \bibliography{ionTransport.bib}


\end{document}